# THE LEGEND OF COSMOLOGICAL HOMOGENEITY

## (How Paradigms Affect Science)

Robert L. Oldershaw

Amherst College

Amherst, MA 01002

rloldershaw@amherst.edu

http://www.amherst.edu/~rloldershaw

**Abstract:** For more than a half century cosmologists have been guided by the assumption that matter is distributed homogeneously on sufficiently large scales. On the other hand, observations have consistently yielded evidence for inhomogeneity in the distribution of matter right up to the limits of most surveys. The apparent paradox can be understood in terms of the role that paradigms play in the evolution of science.

## I. An Assumption Is Born

Modern cosmology dates from the early decades of the 20th century when Einstein's field equations of General Relativity were first applied to the observable universe as a whole. Severe difficulties in solving these equations (10 coupled nonlinear partial differential equations in 4 variables) led theorists to introduce various simplifying assumptions such as the postulate of cosmological homogeneity in order to make the mathematics more manageable. The idea of a uniform cosmos did not represent a break from past assumptions since the previous Newtonian paradigm hypothesized an infinite homogeneous universe. Unexpectedly, however, the global cosmological solutions of Einstein's equations seemed to mandate expanding or contracting universes, and even Einstein himself at first thought that these solutions were probably unrealistic. Subsequently, the work of Edwin Hubble and colleagues demonstrated a roughly linear correlation between the radial velocities and distances of galaxies, and the astrophysical community gradually became convinced that the observable portion of the universe did appear to be undergoing fairly uniform expansion.

At this point the assumption of cosmological homogeneity was still speculative since it could not be thoroughly tested. However, the rough concordance between the theoretical expanding universe model and Hubble's preliminary observations of galactic recession gave astronomers confidence in their new concept of homogeneous expansion. Virtually all astronomers believed that future observations would reveal a highly uniform large-scale distribution of matter. Other problems such as the initial state of the observable universe and the physical details of how the early universe evolved were more pressing concerns. Gradually the Big Bang model of the universe took shape and began to dominate the field of cosmology. This paradigm asserts that approximately 11-15 billion years ago all space, time and matter were packed inside a mysterious entity called a singularity. A cosmological singularity is a mathematical "point" with no size at all, but having infinite density and temperature. If you have trouble imagining such an initial state, do not worry, it is beyond any realistic mental imaging. For unknown reasons the observable universe was believed to have begun expanding from this completely collapsed state. An important point is that the expansion was not envisioned as an explosion taking place *within* spacetime, but rather more like the *creation and expansion of spacetime itself*.

Once the temperature of the adiabatically cooling matter became low enough, structures such as atoms, stars and galaxies could begin to gravitationally "condense" out of the high energy plasma. In the standard Big Bang model it is natural to expect that the uniform expansion would result in homogeneous distributions of background radiations and large-scale structures, so the simplifying assumption of homogeneity seemed quite secure. The envisioned cosmological homogeneity is not expected to be a virtually perfect homogeneity, such as might be found in a pure quartz crystal, but rather it is predicted to be a *statistical* homogeneity which becomes more perfect as you increase the size of the observed volume. In astrophysics, one basic way to test for the presence of this form of statistical homogeneity is to check one vast volume for a uniform distribution of matter, radiation and motion. A slight variation on this method is to select two smaller volumes of space and compare various characteristics of each sample, such as the number of a given class of objects.

In this essay we will see how the assumption of cosmological homogeneity underwent a metamorphosis from a simplifying assumption to an "empirical fact", and how theoreticians have responded to the challenges posed by observations that conflict with the fundamental idea of large-scale homogeneity. This story turns out to be an interesting, surprising and instructive case study for Thomas S. Kuhn's well-known ideas about paradigmatic change and the evolution of science.

II. Troublesome Observations

The concept of cosmological homogeneity started out as a reasonable guess about the universe, and as a mathematical shortcut for making Einstein's equations more tractable. A few decades later the physics community was treating cosmological homogeneity as an observationally verified fact. Since cosmology is now recognized as a full-fledged science, one would expect that this transformation from assumption to apparent fact would be strongly supported by several types of empirical evidence. The curious thing, however, is that unambiguous evidence for cosmological homogeneity has never actually existed. Over the years there were *preliminary* observations of galaxy distributions that were interpreted as supporting the homogeneity assumption. However, more refined observations have shown that such conclusions were usually premature and/or incorrect. It is true that the microwave background radiation, which is interpreted as a primordial remnant of the matter formation era, is cited as very strong support for the cosmological homogeneity assumption. However there are other theoretical explanations for the origin and evolution of this radiation. Moreover, the microwave background has a decisive dipole anisotropy, as well as curious and correlated quadrupole and octopole anomalies, which *may* conflict with using this enigmatic radiation as hard evidence for global homogeneity. For example, some astrophysicists have argued that the observational evidence for large-scale galactic streaming indicates that there is an *intrinsic* dipole structure to the observable universe, rather than a dipole anisotropy induced by a Doppler effect.

Most importantly, the microwave background argument for homogeneity is contradicted by the more direct evidence for *structural inhomogeneity* over a very wide range of observational scales. The microwave background radiation may one day be powerful evidence for or against cosmological homogeneity, but for now it is safer to focus on the distribution of actual galactic scale structures. As our ability to test the homogeneity assumption has improved, a consistent string of discoveries has favored cosmological inhomogeneity right up to the size limits of the large-scale surveys. Here we will look at some of the evidence that has been grudgingly acknowledged in word, but effectively ignored in practice, by many cosmologists.

The first serious challenge to the assumption of cosmological homogeneity came in the latter half of the 1920s when astronomers confirmed that stars were not homogeneously distributed, but rather were amassed in vast "island universes" which we now call galaxies. It was subsequently assumed that beyond the scale of individual galaxies, perhaps at about 1 to 5 Mpc (where 1 Megaparsec = 3.26 x $10^6$ light years), the large-scale distribution of galaxies became homogeneous. However it was not long before astronomers discovered that galaxies were inhomogeneously clustered into small groups of a few to several tens of galaxies. Theoreticians yet again followed the previous pattern: they assumed that the small galaxy clusters would be homogeneously distributed at scales of 20 to 30 Mpc.

As telescopes and data analysis improved, some observers believed that they were seeing inhomogeneous clustering well beyond the previous 30 Mpc limit, perhaps up to 50 Mpc. As a result, the debate over homogeneity versus inhomogeneity began to heat up during the 1970s. In a remarkable article that heralded much of what was to come ("The Case for a Hierarchical Cosmology", *Science* vol. **167**, 1203, 1970), the late Gerard de Vaucouleurs of the University of Texas at Austin presented observational and theoretical evidence indicating that small galaxy clusters were further clustered into "superclusters" on scales of 60 Mpc or greater. In virtually a lone voice of dissent he argued compellingly that there had never been valid evidence for the "empirical fact" of homogeneity. Most theoreticians and the majority of all astronomers at that time disagreed with de Vaucouleurs and believed that inhomogeneity declined sharply at 20 Mpc, with homogeneity taking over just beyond that size scale. The lopsided debate over the reality of superclustering and inhomogeneity at roughly 60 Mpc continued throughout the 1970s, with de Vaucouleurs and a minority of supporters eventually being vindicated.

Not surprisingly, proponents of cosmological homogeneity then predicted that large scale uniformity would finally be found on scales of 60 to 100 Mpc. Once again, however, nature was to disappoint them. During the 1980s the observational evidence for nonuniformity at ever larger scales began to clamor for recognition and explanation. Technical advances led to larger and more detailed surveys of galaxy distributions and motions, and it became clear that inhomogeneity did not "decline sharply beyond 20 Mpc", but rather continued right up to the new observational limits. Even more surprising was the observation that galaxies appeared to be gathered into immense sheets and filaments surrounding bubble-like "voids", wherein the galaxy density was very low. The large scale structure had a mysterious frothy or cellular configuration that is reminiscent of very high energy plasmas, with galaxies playing the role of charged plasma "particles".

By means of a new type of survey that measured the non-radial velocity vectors of galaxies (after general expansion had been subtracted out), it was found that galaxies within 65 Mpc, and later within 100 to 200 Mpc, were flowing in a coordinated manner in a particular direction. These bulk flows suggested that vast lumps of matter on even larger scales were responsible for directing these galactic flows. The most famous lump was dubbed "The Great Atractor".

Meanwhile observers continued to chalk up a steady string of "largest" galaxy superclusters (200 Mpc, 300 Mpc, 360 Mpc, 400 Mpc, …) and voids (100 Mpc, 200 Mpc, 300 Mpc, …).

There were further observational discoveries that shocked nearly everybody.  In one study published in *Nature* by P. Birch of the University of Manchester's Nuffield Radio Astronomy Laboratory at Jodrell Bank, classical double radio galaxies were reported to have nonrandom position angles and polarization vectors on scales of 1,000 Mpc, which was far beyond the scale at which "random" behavior was supposed to have taken over.  Another study, also published in *Nature*, by T.J. Broadhurst *et al* of the University of Durham, revealed evidence for periodic clustering of galaxies over a length scale of 2,000 Mpc!  Both of these reports initially caused a flurry of excitement, with astronomers saying that the results were completely unexpected and required a thorough review of previous assumptions .  However, even though subsequent observations supported the initial findings, most theoretical astrophysicists and cosmologists reverted to their standard assumptions as soon as it seemed respectable to do so.

Toward the end of the 1980s the homogeneity/inhomogeneity debate had become a more even-handed affair, and the sides were more polarized than ever.  Princeton's P.J.E. Peebles, usually one of the more moderate spokesmen for the theoretical cosmology community, stated in *Physica D* (vol. **38**, 273, 1989): "I think the evidence for [large-scale homogeneity] is close to compelling, though it is fair to say, not definitive."  Observational astronomers were more inclined to agree with the conclusions of B.R. Tully of the University of Hawaii (*Science*, vol. **238**, 894, 1987): "A decade ago, we'd have thought that as we went to larger scales we'd see more homogeneity in the universe.  In fact, we see more inhomogeneity."  Those who saw evidence for inhomogeneity right up to the largest scales surveyed tended to think that we really did not understand the large-scale universe very well at all.  Theoreticians, on the other hand, had a resilient faith in their homogeneous models and firmly believed that they nearly had the universe figured out.  Such radically differing interpretations of the same data naturally led to intellectual tension in the astrophysical community.

During the 1990s, the patterns established in the 1980s continued: inhomogeneity was discovered on ever-larger scales and the hypothesis of homogeneity continued its strategic retreat.  The reality of structures and voids in the 100 Mpc to 400 Mpc range was verified.  The periodic clustering over scales on the order of $10^3$ Mpc discovered by Broadhurst *et al* was backed up by subsequent observations.  The "Great Attractor" was dethroned by the even larger "Great Wall", a vast and roughly linear agglomeration of galaxy clusters in the northern hemisphere.  Then a surprisingly similar southern hemispheric counterpart dubbed the "Southern Wall" was discovered.  The evidence for large-scale inhomogeneity was reviewed perceptively by P.H. Coleman of the University of Leiden and L. Pietronero of the University of Rome (*Physics Reports*, vol. **213**, 311, 1992).  They argued that the proposed evidence in favor of cosmological homogeneity was "based on methods of analysis that assume it implicitly."  A more objective analysis of the data convinced them that the evidence for nonuniformity persisted right up to the limits of all surveys, such that there was no credible evidence for convergence to homogeneous distributions.  Their analysis suggested a fractal distribution of matter extending up to the observational limits, as de Vaucouleurs had presciently envisioned two decades before.  Over the next 16 years the arguments for fractal distributions of matter on very large scales increased in number and strength.  By 2008 Pietronero and his colleagues F. Sylos Labini, N.L. Vasilyev and Y.V. Baryshev felt that their analysis of the Sloan Digital Sky Survey data showed conclusively that fractal structure persisted up to the new observational limits, and that their results were incompatible with cosmological homogeneity on scales lower than 100 Mpc.

Throughout this time period the responses of those who have long favored homogeneous models of the universe have been quite interesting. They have confidently asserted in books and in scientific papers that the universe simply *must be homogeneous*, if not at 60 Mpc or at 400 Mpc then surely *somewhere* beyond that. One might infer that these scientists do not attach a great deal of importance to the observational evidence for cosmological inhomogeneity that has been accruing over the last few decades. We have here a somewhat unusual state of affairs with two groups of describing the same observable universe in two very different ways. One group seriously thinks that we are on the verge of a nearly complete understanding of the cosmos and speaks of having entered the era of "precision cosmology." The other group worries that some of the key cosmological assumptions that we have embraced for decades might be misleading us, and that our understanding of the cosmos might actually be quite limited and rudimentary.

### III. Battle of the Paradigms

How is it possible that these two groups can arrive at such mutually exclusive conclusions? Actually, this apparently paradoxical state of affairs regarding cosmological homogeneity is not all that hard to understand. When humans repeatedly use an assumption or model, it often gradually becomes transformed into "fact." In perhaps the best known treatise on the evolution of scientific theories, *The Structure of Scientific Revolutions*, T.S. Kuhn documents how scientists become so steeped in the prevailing paradigm that healthy skepticism about its underlying assumptions tends to disappear. What begins as creative speculation metamorphoses into "common sense" and self-evident truth, doubted only by 'the great unwashed' and 'wrong-headed' mavericks. Usually the critical assumptions are supported by tentative evidence that, with all good intentions, is subtly shaped to fit theoretical needs. On the other hand, contradictory evidence is all too often ignored or belittled. For example, one cosmologist referred to recent evidence solidly lining up behind cosmological inhomogeneity as "anecdotal." As the reigning paradigm becomes increasingly entrenched, many scientists tend to look for and, not surprisingly, find what they *expect* to find. It is far more difficult and counterintuitive to recognize the importance of confusing and "unwanted" results that are not expected.

When a new scientific paradigm begins to mount a serious challenge to the reigning paradigm, we see the development of an impasse such as the one discussed above, where two groups use the same general body of evidence to defend competing and mutually exclusive world views. It is important to recognize that paradigmatic consensus is *usually* beneficial to scientific progress because it organizes knowledge, helps in generating new questions and coordinates scientific activities. Unfortunately, however, there is a problematic side to the consensus approach. Scientists who have studied, conducted research and made their mark under the guidance of a prevailing paradigm are often confined to its conceptual limits. Competing paradigms sound decidedly wrong to them, since the new hypotheses and assumptions seem to be self-evident violations of hard-won scientific knowledge. Moreover, competing paradigms represent a threat to one's professional status and one's sense of intellectual security. The transition from an old paradigm to a new one requires scientists to give up deeply held beliefs and to accept new ones that sound somewhat outrageous. Obviously this is not something that comes easily to scientists or laypersons.

The debate over the legend of cosmological homogeneity may be symptomatic of an ongoing, slow-motion revolution in cosmology. Major contests between old and new paradigms are not usually fought and won quickly, but rather are more like the collision of two glaciers. The more massive one will eventually overpower the other, but the action is slow-paced and at times it is not entirely clear which side has the momentum in its favor. Will the theoretician's dream of a homogeneous cosmos finally emerge before the survey limits reach to the edges of the observable universe, or will the 50-year trend of discovering inhomogeneity on ever-larger scales continue unabated? Do we know nearly everything about a homogeneous universe, or are we still groping for a basic understanding of a fractal universe? Time will tell.